\documentclass{elsart}
\usepackage{graphicx}
\usepackage{amsmath, amssymb}
\allowdisplaybreaks[4]
\newcommand{\vect}[1] {\mathbf{#1}}
\newcommand{\dif} {\mathrm{d}}
\newcommand{\up} {\uparrow}
\newcommand{\down} {\downarrow}

\newcommand{\cI}{\mathcal{I}}

\begin{document}

\begin{frontmatter}
\title{Generalized Virial Theorem and Pressure Relation for a strongly correlated Fermi gas}
\author{Shina Tan}
\address{Institute for Nuclear Theory, University of Washington, Seattle, WA 98195-1550, USA}

\begin{abstract}
For a two-component Fermi gas in the unitarity limit (ie, with infinite scattering length), there is a well-known
virial theorem, first shown by J. E. Thomas et al, Phys. Rev. Lett. 95, 120402 (2005). A few people rederived this result, and extended it to few-body systems, but their results are all restricted to the unitarity limit.
Here I show that there is a generalized virial theorem for FINITE scattering lengths.
I also generalize an exact result concerning the pressure, first shown in cond-mat/0508320, to the case of imbalanced populations.
\end{abstract}

\end{frontmatter}

Two-component ultracold atomic Fermi gases with large scattering lengths have been realized recently, and have become a focus of
numerous research activities. In this paper we study two exact properties of such a system.

We will consider the zero-range interaction model only, in which the scattering length $a$ between the $\up$ and $\down$ spin states is
the only parameter for the interaction. (Such a model is justified by typical experimental
setups, in which the interatomic distance, the thermal de Broglie wavelength, and $a$ are all large compared to the
Van de Waals range of the interaction.)

\section{Generalized Virial Theorem}
If the system is confined by a harmonic trap, and is in the unitarity limit ($a\rightarrow\infty$),
the total energy $E$ is related to the external potential energy by
\begin{equation}\label{eq:special_virial}
E=2E_V.
\end{equation}

The result \eqref{eq:special_virial} was first shown by J. E. Thomas et al using the local density approximation \cite{Thomas2005PRL}.
A few people rederived this result, and extended it to few-body systems, but their results are all restricted to the unitarity limit and to a harmonic confinement potential \cite{SpecialVirialTheorem_others}.

Here I show that there is a \emph{generalized} virial theorem for \textit{finite} scattering lengths. I also consider a somewhat more
general confinement potential, 
$$V(\vect r)=r^\beta f(\hat{\vect r})$$
satisfying $\beta>-2$, $\beta\ne0$, and $\beta f(\hat{\vect r})>0$,
where $f(\hat{\vect r})$ is any smooth function of the unit direction vector $\hat{\vect r}$.
For a harmonic trap $\beta=2$.

Such a generalized virial theorem is
\begin{equation}\label{eq:general_virial}
E-\frac{\beta+2}{2}E_V=-\frac{\hbar^2\cI}{8\pi am},
\end{equation}
where $m$ is the fermion mass,
$\cI=\lim_{k\rightarrow\infty} k^4\rho_{\vect k\sigma}$, and $\rho_{\vect k\sigma}$ is the momentum distribution
at momentum $\hbar\vect k$ and spin state $\sigma$.
The amplitude of $\rho_{\vect k\sigma}$ is defined by $\int\frac{\dif^3k}{(2\pi)^3}\rho_{\vect k\sigma}=N_\sigma$,
the number of spin-$\sigma$ fermions. Note that $N_\up$ and $N_\down$ are arbitrary and may be different.

$\cI$ equals the quantity $\Omega C$ in Refs.~\cite{ShinaTan0505200,ShinaTan0508320}.
An equivalent definition of $\cI$ is given in Ref.~\cite{ShinaTan0508320}:
$$\cI=\lim_{K\rightarrow\infty}\pi^2KN_{k>K},$$
where $N_{k>K}$ is the expectation of the total number of fermions with momenta larger than $\hbar K$.

We will give $\cI$ a short name: \emph{total contact}, or simply \emph{contact}.
We will give the spatial function $C(\vect r)$ introduced in Ref.~\cite{ShinaTan0505200} a name:
\emph{local contact density}. The quantity $C=\Omega^{-1}\cI=\Omega^{-1}\int\dif^3 rC(\vect r)$ \cite{ShinaTan0505200}
will be called \emph{average contact density} (over volume $\Omega$). For a homogeneous system of volume $\Omega$,
$C(\vect r)$ equals $C$.

To prove \eqref{eq:general_virial}, we first consider an energy eigenstate $\phi$ at scattering length $a$,
with energy $E=E_\text{internal}+E_V$,where $E_\text{internal}$ is the internal energy expectation value.
We then modify this state infinitesimally, in two consecutive steps.

In the first step, we adiabatically change the scattering length to $a'=(1+\epsilon)a$, where $\epsilon$ is an infinitesimal number.
The energy changes to $E'=E'_\text{internal}+E'_V$.
Using the adiabatic sweep theorem of Ref.~\cite{ShinaTan0508320}, we find
$$E'-E=\frac{\hbar^2\cI}{4\pi am}\epsilon+O(\epsilon^2).$$

In the second step, we do a geometric compression of the system's wave function, from $\phi'(\vect r_1,\cdots,\vect r_N)$
to $$\phi''(\vect r_1,\cdots,\vect r_N)=(1+\epsilon)^{3N/2}\phi'\mathbf{(}(1+\epsilon)\vect r_1,\cdots,(1+\epsilon)\vect r_N\mathbf).$$
Using the short-range boundary condition for the wave function ($\phi\propto 1/s-1/\text{scatt.length}$
when the distance $s$ between two fermions
in different spin states is small), we find that $\phi''$ corresponds to a state with scattering length
$a''=a'/(1+\epsilon)=a$. Using the energy theorem \cite{ShinaTan0505200}, we get 
$E''_\text{internal}=(1+\epsilon)^2E'_\text{internal}$, and $E''_V=(1+\epsilon)^{-\beta}E'_V$.
So $$E''-E'=2\epsilon E'_\text{internal}-\beta \epsilon E'_V+O(\epsilon^2)=2\epsilon E_\text{internal}-\beta \epsilon E_V+O(\epsilon^2).$$

Because the state $\phi''$ has the \emph{same} scattering length as the initial state $\phi$,
and because the difference between the two states is of the order $\epsilon$, the variational stability of energy levels implies that
$E''-E=O(\epsilon^2)$, or $(E''-E')+(E'-E)=O(\epsilon^2)$. So
$$\frac{\hbar^2\cI}{4\pi am}+2 E_\text{internal}-\beta E_V=0.$$
Rewriting $E_\text{internal}=E-E_V$, we get \eqref{eq:general_virial}.

Obviously, \eqref{eq:general_virial} is also valid for any statistical ensemble of energy levels, with a statistical weight
decaying sufficiently fast at large energy, such that $\cI$ equals the statistical average of the values of $\cI$'s
for the individual energy levels \cite{ShinaTan0505200}. Thus \eqref{eq:general_virial} is valid for any finite temperature states
in the canonical or grand canonical ensemble, as well as the ground state.

When $k_Fa\rightarrow0^-$ ($k_F$ fixed), $\cI\propto a^2$ and Eq.~\eqref{eq:general_virial} reduces to the virial theorem
for the noninteracting Fermi gas.

When $a=\infty$ and $\beta=2$, Eq.~\eqref{eq:general_virial} reduces to Eq.~\eqref{eq:special_virial}.

When $N_\up=N_\down\gg1$, $k_Fa\rightarrow0^+$ and the temperature is zero, the system
approaches a Bose-Einstein condensate
of tightly bound molecules, and \eqref{eq:general_virial} approaches the virial theorem for the Gross-Pitaevskii equation
for these bosonic molecules \cite{virial_theorem_BEC} with scattering length $a_m\approx0.6a$ \cite{Petrov2004PRL}.

\section{Pressure Relation}
Suppose that the system is in a cubic box of size $L=\Omega^{1/3}$, and a periodic boundary condition is imposed.
In the absence of the external potential
\begin{equation}\label{eq:P}
P-\frac{2}{3}\rho_E=\frac{\hbar^2C}{12\pi am},
\end{equation}
where $\rho_E=E/\Omega$ is the average energy density, and $C=\cI/\Omega$ is the average contact density.
Equation~\eqref{eq:P} is valid for any energy eigenstate, or any statistical ensemble of them, with a statistical weight
decaying sufficiently fast at large energy, such that $\cI$ equals the statistical average of the values of $\cI$'s
for the individual energy levels \cite{ShinaTan0505200}. This includes any finite temperature states
in the canonical or grand canonical ensemble, as well as the ground state.

In Ref.~\cite{ShinaTan0508320}, the pressure relation \eqref{eq:P}
is shown for balanced populations of the two spin states: $N_\up=N_\down$.

Here I point out that \eqref{eq:P} remains valid \emph{even if $N_\up\ne N_\down$}. 
This incorporates many interesting possibilities, in particular phase separation between superfluid and normal
phase \cite{phase_separation} and, consequently, spontaneous spatial inhomogeneity of the energy density and contact density.

The general proof of \eqref{eq:P} is very similar to that of \eqref{eq:general_virial}.
Starting with any energy eigenstate $\phi$ with scattering length $a$ and energy $E$,
we first increase the scattering length adiabatically, from $a$ to $a'=(1+\epsilon)a$, to get 
$E'=E+\frac{\hbar^2\cI}{4\pi am}\epsilon+O(\epsilon^2)$.
We then do a geometric compression of the wave function, after which the scattering length changes back to $a$,
the period of the wave function becomes $L/(1+\epsilon)$, the energy becomes 
$$E''=(1+\epsilon)^2E'=E'+2\epsilon E'+O(\epsilon^2)=E+\frac{\hbar^2\cI}{4\pi am}\epsilon+2\epsilon E+O(\epsilon^2),$$
and the quantum state becomes $\phi''$.

If we start from the state $\phi$, and compress the box by a linear factor $(1+\epsilon)$ adiabatically, without changing
the scattering length, we will get the \emph{same} final state as $\phi''$. So the pressure is
$$P=\lim_{\epsilon\rightarrow0}\frac{E''-E}{\Omega-(1+\epsilon)^{-3}\Omega}=\frac{2E}{3\Omega}+\frac{\hbar^2\cI}{12\pi am\Omega}.$$

Although Eq.~\eqref{eq:P} is only exact in the absence of external potential, it is approximately valid for each local part of the fermionic
cloud in a trap, within the local density approximation. In this latter case, $\rho_E$ is replaced by the local internal energy density,
and $C$ is replaced by the local contact density $C(\vect r)$.

\ack
The author thanks L.~I.~Glazman and K.~Levin for communications, and thanks E.~Braaten for suggesting
intuitive names for the quantities $\cI$ and $C(\vect r)$. This work was supported by DOE Grant
No. DE-FG02-00ER41132.

\bibliographystyle{elsart-num}
\bibliography{virial}

\end{document}